\documentclass[10pt,pre,twocolumn,byrevtex,superscriptaddress,showpacs]{revtex4}
\usepackage{graphics}

\begin{document}

\title{Fluctuation relation for a L\'evy particle}

\author{H. Touchette}
\email{ht@maths.qmul.ac.uk}
\affiliation{\mbox{School of Mathematical Sciences, Queen Mary, University of London, London E1 4NS, UK}}

\author{E. G. D. Cohen}
\affiliation{The Rockefeller University, 1230 York Avenue, New York, New York 10021, USA}

\date{\today}

\begin{abstract}
We study the work fluctuations of a particle subjected to a deterministic drag force plus a random forcing whose statistics is of the L\'evy type. In the stationary regime, the probability density of the work is found to have ``fat'' power-law tails which assign a relatively high probability to large fluctuations compared with the case where the random forcing is Gaussian. These tails lead to a strong violation of existing fluctuation theorems, as the ratio of the probabilities of positive and negative work fluctuations of equal magnitude behaves in a non-monotonic way. Possible experiments that could probe these features are proposed. 
\end{abstract}


\maketitle

The study of fluctuations in systems driven in nonequilibrium steady states has concentrated lately on a special symmetry property referred to as the fluctuation theorem or fluctuation relation. The first studies of this symmetry focused on the entropy production of nonequilibrium systems \cite{evans1993,gallavotti1995}, but it was soon realized that the same symmetry holds for other quantities of interest, such as the work $W_\tau$ performed on a nonequilibrium system during a time $\tau$ \cite{cohen2003}. In the context of this quantity, we define a fluctuation relation \cite{note1} as follows. Assuming that $W_\tau$ is extensive in $\tau$, we say that the probability density $P(W_\tau)$ of $W_\tau$ satisfies a fluctuation relation or, to be more precise, a \emph{conventional} fluctuation relation if   
\begin{equation}
\frac{P(W_\tau=w \tau )}{P(W_\tau=-w\tau)} = e^{c\tau w},
\label{ft1}
\end{equation}
where $c$ is a constant which depends neither on $\tau$ nor $w$. We also say that $P(W_\tau)$ satisfies a conventional fluctuation relation if the above equality holds, when properly scaled (see below), in the limit $\tau\rightarrow\infty$. In that case, we speak of a \textit{stationary} conventional fluctuation relation.

One important problem about fluctuation relations, which is as yet unresolved, is to determine the class of nonequilibrium systems or, more precisely, the class of nonequilibrium observables \cite{note2} whose fluctuations satisfy a conventional fluctuation relation. Our understanding of this issue at this point is that this relation holds for two general classes of systems: (i) finite, deterministic systems having a sufficiently chaotic dynamics \cite{gallavotti1995,gallavotti1995b}, and (ii) finite, stochastic systems whose evolution is a Markov process \cite{kurchan1998}. There are subtle points to be taken into account, however. In some cases, boundary conditions or special forms of large deviations may restrict the range of validity of conventional fluctuation relations, and in these cases, corrections or extensions of the conventional fluctuation relation have been proposed \cite{cohen2003b,farago2002}. 

Our goal in this paper is to expand this picture of fluctuation relations by studying a model of a nonequilibrium system whose work fluctuations neither obey the conventional fluctuation relation nor the extended fluctuation relation of van Zon and Cohen \cite{cohen2003b}. Based on this model, we propose a novel type of fluctuation relation, and compare it, from the general point of view of large deviation theory, with the conventional fluctuation relation defined in (\ref{ft1}). In the end, we also discuss two experiments for which ``anomalous'' fluctuations similar to the ones found for our model could be observed. The first experiment is related to dry friction, while the second draws on recent experiments on granular gases.

The specific model that we study is a variation of the Langevin model studied recently by van Zon and Cohen \cite{cohen2003b}, which describes the motion of a particle subjected to three forces: a deterministic force derived from a particle-confining harmonic potential, a friction force, and a random force, i.e., a noise. The combined effect of these forces on the particle is described, in the overdamped limit, by the Langevin equation
\begin{equation}
\dot{x}(t)=-\gamma [x(t)-x^{*}(t)]+\xi (t),
\end{equation}
where $x(t)$ denotes the position of the particle at time $t$, $x^{*}(t)$ the position of the center of the harmonic potential at $t$, $\gamma$ a constant related to the ratio of the strengths of the potential and the friction, and $\xi(t)$ the random forcing. For simplicity, we assume that the center of the potential moves at constant speed $v^*$, i.e., $x^{*}(t)=v^{*}t$. Finally, and this is where we depart from the model of \cite{cohen2003b}, we assume that $\xi(t)$ represents a L\'evy white noise, i.e., an uncorrelated noise obeying a L\'evy statistics \cite{west1982}. Such a noise process is entirely defined by its characteristic function 
\begin{equation}
G_\xi [k]=\left\langle e^{ik\cdot \xi }\right\rangle =\int \mathcal{D}[\xi
]\ P[\xi ]\ e^{i\int_0^\infty k(t)\xi (t)dt},
\end{equation}
which, in the case of symmetric L\'evy white noise, is taken to have the form \cite{west1982,uchaikin1999}
\begin{equation}
G_\xi [k]=\exp \left( -b\int_0^\infty |k(t)|^\alpha dt\right)
\label{cf1}
\end{equation}
with $b>0$ and $\ 0<\alpha \leq 2$. Gaussian white noise is a special case of the noise process, obtained by choosing $\alpha=2$. For this case, which is the typical case considered by van Zon and Cohen, $\langle \xi (t)\rangle =0$ and $\langle \xi (t)\xi (t^{\prime })\rangle =2 b \delta (t-t^{\prime })$.

The introduction of L\'evy noise with $\alpha\in(0,2)$ in Langevin equations is often perceived as being  physically unreasonable because it causes $x(t)$ to have an infinite variance, which implies that there can be no fluctuation-dissipation relation connecting the strength of the friction to the variance of the noise \cite{west1982}. This criticism indeed applies if we regard the noise as being \textit{internal}, i.e., as arising, like the friction, from the environment surrounding the particle. However, if we regard the noise as being \textit{external}, i.e., as being an exogenous driving force, then there is no problem in assuming that $\xi(t)$ is L\'evy-distributed. In that case, the friction and the external noise are decoupled, and one is free to adopt any model of noise. This point will be clarified again when we discuss the experiments that we have alluded to above. 

For now we leave these considerations aside, and attempt to calculate the probability density $P(W_\tau)$ of the work $W_\tau$ performed on the L\'evy particle by the harmonic drag force during a time $\tau$ \cite{cohen2003}:
\begin{equation}
W_\tau =-v^{*}\int_0^\tau [x(t)-x^{*}(t)]dt.
\label{defw1}
\end{equation}
To carry out this calculation, it is convenient to study the dynamics of $x(t)$ in a comoving frame with the change of coordinate $y(t)=x(t)-x^{*}(t)$, so that $W_\tau=-v^{*}\int_0^\tau y(t)dt$. The dynamical equation for $y(t)$ reads 
\begin{equation}
\dot{y}(t)=-\gamma y(t)+\eta (t),
\label{lang2}
\end{equation}
where $\eta (t)=\xi (t)-v^{*}$ represents a white noise with mean or ``drift'' $-v^*$.

At this point, we proceed to calculate $P(W_\tau )$ by calculating its characteristic function, defined as 
\begin{equation}
G_{W_\tau }(q)=\left\langle e^{iqW_\tau }\right\rangle =\int_{-\infty
}^\infty P(W_\tau=w)\ e^{iqw}dw.
\end{equation}
A key mathematical observation here is that this characteristic function can be obtained from the characteristic function $G_y[k]$ of the process $y(t)$ by using $k(t)=-v^* q$ for $t\in[0,\tau]$ and $k(t)=0$ otherwise. With this choice of $k(t)$, we indeed obtain from Eqs.~(\ref{cf1}) and (\ref{defw1}) 
\begin{equation}
G_y[k] =\left\langle \exp \left( i\int_0^\infty k(t)y(t)dt\right)
\right\rangle  =G_{W_\tau }(q).
\end{equation}

In order to find $G_y[k]$, we next solve the C\'aceres-Budini formula associated with $y(t)$ \cite{caceres1997}:
\begin{equation}
G_y[k] = G_\eta [r],\quad r(l) =\int_l^\infty e^{\gamma (l-t)}k(t)dt.
\label{cbeq1}
\end{equation}
In applying this formula to our model, we assume that $y(0)=x(0)=0$. We also adopt at this point dimensionless units by setting $\gamma=1$ and $b=1$. Noticing that
\begin{equation}
G_\eta [r] =\exp \left( -iv^{*}\int_0^\infty r(l)dl\right) G_\xi [r],
\end{equation}
we can write
\begin{equation}
G_{W_\tau }(q)=\exp \left( -iv^{*}\int_0^\infty r(l)dl-\int_0^\infty |r(l)|^\alpha dl\right).  
\label{ceq1}
\end{equation}
By performing the integrals involved in this expression, we arrive at
\begin{equation}
G_{W_\tau }(q)=e^{iM q-V |q|^\alpha},
\end{equation}
where
\begin{eqnarray}
M &=& \nu (\tau +e^{-\tau }-1) \label{m1}\\
V  &=& \nu^{\alpha /2} \int_0^\tau |e^{l-\tau}-1|^\alpha\ dl,
\end{eqnarray}
having defined $\nu=(v^*)^2$. 

\begin{figure}[t]
\resizebox{3.4in}{!}{\includegraphics{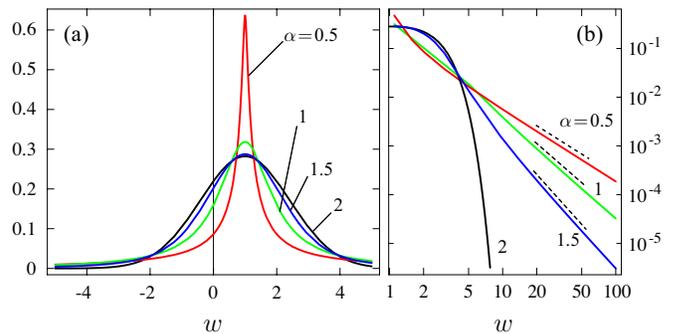}}
\caption{(Color online) (a) Plot of $P(W_\tau=w)$ for various values of $\alpha$ ($M=V=1$). (b) Log-Log plot of $P(W_\tau=w)$.}
\label{figdens1}
\end{figure}

This characteristic function is the main mathematical result of this paper, from which $P(W_\tau)$ is obtained by an inverse Fourier transform. In the present case, there is no need to explicitly compute the inverse Fourier transform because the form of $G_{W_\tau}(q)$ shown in (\ref{ceq1}) already tells us that $P(W_\tau)$ is a symmetric L\'evy distribution having the same index $\alpha$ as the noise $\xi(t)$ \cite{uchaikin1999}. The parameter $M$ is the value at which $P(W_\tau)$ is centered, while $V$ is a measure of the width of $P(W_\tau)$; see Fig.~\ref{figdens1}. The index $\alpha$, finally, determines how the tails of $P(W_\tau=w)$ decay as $|w|\rightarrow\infty$. It is at this point that we encounter two very different types of fluctuations.

A first type arises from Gaussian white noise, which is obtained again with $\alpha=2$. For this case, and for the particular initial condition that we consider here, namely $x(0)=0$, $P(W_\tau)$ has an exact Gaussian form
\begin{equation}
P(W_\tau=w)=\frac{1}{\sqrt{2\pi \sigma^2}} e^{-(w-M)^2/(2\sigma^2)}
\end{equation}
with mean $M$ given by (\ref{m1}) and variance $\sigma^2$ given by
\begin{equation}
\sigma^2 = 2V= \nu [2\tau-3+4 e^{-\tau} -e^{-2\tau}].
\end{equation}
This form of $P(W_\tau)$ obeys a stationary conventional fluctuation relation. Indeed, since $M\sim \nu\tau$ and $\sigma^2\sim 2\nu\tau\sim 2M$ in the limit $\tau\rightarrow\infty$, we have\begin{equation}
g_\tau(p)=\frac{P(W_\tau=p\nu\tau)}{P(W_\tau=-p\nu\tau)}\approx e^{p\nu\tau}
\label{ft2}
\end{equation}
with sub-exponential corrections in $\tau$. This result confirms what van Zon and Cohen have reported in \cite{cohen2003b} using a different calculation technique. Here, $p$ is the scaled (intensive) value of $W_\tau$ defined as $p=W_\tau/\langle W_\tau\rangle$, with the brackets denoting an ensemble avarage.

\begin{figure}[t]
\resizebox{3.4in}{!}{\includegraphics{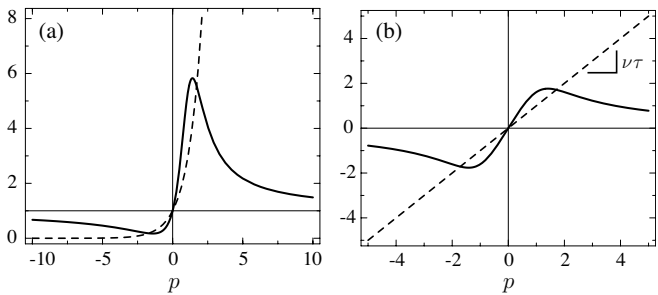}}
\caption{(a) Comparison of $g_\tau(p)$ and (b) $\ln g_\tau(p)$ for the Cauchy (solid line) and Gaussian (dashed line) cases ($\alpha=1$, $\nu=1$, and $\tau=1$ for the Gaussian case).}
\label{figcomp1}
\end{figure}

An entirely different case of fluctuations arises when $\alpha< 2$, i.e., when $\alpha\in(0,2)$. In this case, $P(W_\tau)$ is not Gaussian anymore. Rather, it is a so-called \emph{strict} L\'evy distribution \cite{uchaikin1999} having, as is well known, power-law tails of the form
\begin{equation}
P(W_\tau=w)\sim |w|^{-1-\alpha}
\label{tail1}
\end{equation}
as $w\rightarrow\pm\infty$; see Fig.~\ref{figdens1}(b). Because of these power-law tails, it is clear that $P(W_\tau)$ does not satisfy a fluctuation relation of the form defined by Eq.~(\ref{ft1}), even in the long time (stationary) regime. To be sure, consider the case $\alpha=1$, for which we have the following exact solution:
\begin{equation}
P(W_\tau=w)=\frac{1}{\pi} \frac{V}{(w-M)^2+V^2},
\end{equation}
with $M$ given, as before, by Eq.~(\ref{m1}) and $V=\nu^{1/2}(\tau-1+e^{-\tau})\sim\nu^{1/2}\tau$. This form of distribution is known as a Cauchy or Lorentz distribution, and behaves like $|w|^{-2}$ for large work values. Using this explicit form of $P(W_\tau)$, we obtain in the limit $\tau\rightarrow\infty$
\begin{equation}
g_\tau(p)\sim \frac{\nu(p+1)^2+1}{\nu(p-1)^2+1}.
\label{ft3}
\end{equation} 
As shown in Fig.~\ref{figcomp1}, this form of $g_\tau(p)$ is not a monotonic function of $p$, contrary to its Gaussian counterpart. In the Cauchy case, we further have that $g_\tau(p)$ does not asymptotically depend on time, which implies that measuring $W_\tau$ over longer and longer integration times $\tau$ leads to no appreciable difference in the behavior of the fluctuations of the scaled variable $p$. In the Gaussian case, $g_\tau(p)$ does depend on time, with the consequence that negative fluctuations of the work get exponentially suppressed as $\tau\rightarrow\infty$. 

The asymptotic time independence of $g_\tau(p)$ for Cauchy white noise is special, and does not carry over to other values of $\alpha$ in the range $(0,2)$. However, since all strict L\'evy distributions have power-law tails, $P(W_\tau)$ cannot satisfy a conventional fluctuation relation for any $\alpha\in(0,2)$. Indeed, because of (\ref{tail1}), we have
\begin{equation}
\lim_{p\rightarrow\pm\infty} g_\tau(p)= 1
\label{asymp1}
\end{equation}
for all $\alpha\in(0,2)$ (Fig.~\ref{figcomp1}). In words this means that large negative fluctuations of $W_\tau$ are just as likely to happen as large positive fluctuations of equal magnitude. For $\alpha=2$, by comparison, we have $g_\tau(p)\rightarrow\infty$ when $p\rightarrow\infty$ and $g_\tau(p)\rightarrow 0$ when $p\rightarrow -\infty$, which implies again that positive fluctuations of $W_\tau$ are exponentially more probable than negative fluctuations. This difference in behavior clearly shows that the fluctuations of the work performed on a particle driven by \textit{strict} L\'evy white noise satisfy an entirely different kind of fluctuation relation when compared with its Gaussian noise counterpart. Following a terminology used in studies of L\'evy noise \cite{uchaikin1999}, we call this new fluctuation relation, satisfying both (\ref{tail1}) and (\ref{asymp1}), an \textit{anomalous} fluctuation relation.

An important property of the density $P(W_\tau)$ brought about by its power-law tails is that it does not have the form of a large deviation probability. By this we mean that $P(W_\tau)$ does \emph{not} have the asymptotic form
\begin{equation}
P(W_\tau = w\tau) \approx e^{-\tau I(w)}
\label{ldt1}
\end{equation}
in the limit $\tau\rightarrow\infty$, where $I(w)$ is some function which does not depend on $\tau$ \cite{ellis1985}. This observation is crucial because it explains why $P(W_\tau)$ fails to satisfy a \textit{normal} (i.e., conventional or extended) fluctuation relation \emph{even if the Langevin model studied here is Markovian}. The fact is that the Markov property is not a sufficient condition for having a normal fluctuation relation, as is demonstrated by our model. What is essential is that the observable of interest be governed by a probability density having a large deviation form. Note indeed that in order for $P(W_\tau)/P(-W_\tau)$ to be exponential in $\tau$, $P(W_\tau)$ must itself be exponential in $\tau$, as in (\ref{ldt1}). Here, $P(W_\tau)$ has a large deviation form only for $\alpha=2$, which is why this value gives rise to a normal fluctuation relation, while all the other values of $\alpha$ give rise to an anomalous fluctuation relation. A more detailed explanation of this point together with a complete analysis of $P(W_\tau)$ for values of $\alpha$ other than $1$ and $2$ will be given in another paper \cite{note3}.   

To complement the theory presented so far, we now suggest an experiment that could be used to observe the anomalous fluctuation relation displayed by our model. As mentioned before, the type of physical system that we must look for is one in which the noise is introduced externally and can be tailored to mimic L\'evy white noise. One such system is provided by an experiment recently reported by Buguin \textit{et al}.~\cite{buguin2005}, in which a solid object placed on a solid plane is set in motion by rapidly vibrating the plane horizontally. In \cite{degennes2005}, de Gennes considered the possibility of vibrating the plane using Gaussian white noise, and provided various theoretical predictions for this case, based on a Langevin equation containing a dry friction term and a noise term, uncorrelated with the friction term, representing the random motion of the plane. Repeating the experiment of Buguin \textit{et al}.~with Gaussian white noise, one could check whether a normal fluctuation relation for the work $W_\tau$ performed by the plane on the moving object during a time $\tau$ holds. Then, using L\'evy white noise, one could check whether an anomalous fluctuation relation holds for $W_\tau$ \cite{note4}.

A different experimental investigation of anomalous fluctuations could be based on a gas of inelastically-colliding particles, maintained in steady states by rapidly vibrating its container. Several studies of this so-called ``granular gas'' have appeared recently (see, e.g., \cite{aumaitre2001,bonetto2006b}), and focus on the fluctuations of the power injected by the vibrations. In this context, two types of vibrations have been considered, namely periodic forcing and Gaussian white noise, and for these it appears that normal fluctuation relations hold in the short time regime where $\tau$ is comparable with the particles' mean free time \cite{aumaitre2001}. In view of our model, one could change the vibrations from Gaussian to L\'evy white noise, and see whether an anomalous fluctuation relation similar to the one found here holds for the short time fluctuations of the injected power. One could also check that the distribution of the injected power does not have the form of a large deviation probability when strict L\'evy noise is used. This can be done in simulations, and, at least in principle, in real experiments. 

Note that L\'evy noise, like Gaussian noise, is a \textit{proper} form of noise, in the sense that its realizations are always finite \cite{uchaikin1999}. Therefore, forcing an object with a L\'evy noise having an infinite variance, or even an infinite mean, does not imply that we supply that object with an infinite amount of energy. In particular, L\'evy noise cannot supply an infinite amount of heat to a granular gas in a finite amount of time.

To summarize, we have shown that the work fluctuations of a nonequilibrium system can show different behaviors depending on the form of noise perturbing the system. This was illustrated with a Langevin equation perturbed by L\'evy white noise. For this model, we have shown that the probability density of the work fluctuations does not have the form of a large deviation probability, since it has power-law tails. These fluctuations lead therefore to a novel type of fluctuation relation, which differs from the conventional and extended fluctuation relations discussed up until now. Other noises behaving asymptotically like L\'evy white noise should give rise to similar results, since L\'evy distributions are stable \cite{uchaikin1999}. In this sense, the results obtained here should be representative of a large class of fluctuation relations associated with ``power-law'' noises having infinite variances.

\begin{acknowledgments}
This work was supported by NSERC (Canada), the Royal Society of London, HEFCE (England), and the National Science Foundation (PHY-0501315).
\end{acknowledgments}


\begin{thebibliography}{99}
\bibitem{evans1993} D. J. Evans, E. G. D. Cohen, and G. P. Morriss, Phys. Rev. Lett. \textbf{71}, 2401 (1993); D. J. Evans and D. J. Searles, Phys. Rev. E \textbf{50}, 1645 (1994); D. J. Searles and D. J. Evans, J. Chem. Phys. \textbf{113}, 3503 (2000).

\bibitem{gallavotti1995} G. Gallavotti and E. G. D. Cohen, Phys. Rev. Lett. \textbf{74}, 2694 (1995).

\bibitem{cohen2003} R. van Zon and E. G. D. Cohen, Phys. Rev. E \textbf{67}, 046102 (2003).

\bibitem{note1} We prefer to use the term ``fluctuation relation'' over ``fluctuation theorem'' because a fluctuation relation can be observed numerically or experimentally without being proved as a theorem.

\bibitem{note2} By nonequilibrium observable, we mean a random variable defined with respect to a nonequilibrium system in a steady state.

\bibitem{gallavotti1995b} G. Gallavotti and E. G. D. Cohen, J. Stat. Phys. \textbf{80}, 931 (1995); G. Gentile, Forum Math. \textbf{10}, 89 (1998); C. Maes and E. Verbitskiy, Comm. Math. Phys. \textbf{233}, 137 (2003); A. Giuliani, F. Zamponi, and G. Gallavotti, J. Stat. Phys. \textbf{119}, 909 (2005).

\bibitem{kurchan1998} J. Kurchan, J. Phys. A: Math. Gen. \textbf{31}, 3719 (1998); J. L. Lebowitz and H. Spohn, J. Stat. Phys. \textbf{95}, 333 (1999); C. Maes, J. Stat. Phys. \textbf{95}, 367 (1999).

\bibitem{cohen2003b} R. van Zon and E. G. D. Cohen, Phys. Rev. Lett. \textbf{91}, 110601 (2003); Phys. Rev. E \textbf{69}, 056121 (2004).

\bibitem{farago2002} J. Farago, J. Stat. Phys. \textbf{107}, 781 (2002); L. Rey-Bellet and L. E. Thomas, Ann. Inst. Henri Poincar\'e \textbf{3}, 483 (2002); F. Bonetto, G. Gallavotti, A. Giuliani, and F. Zamponi, J. Phys. Stat. \textbf{123}, 39 (2006); R. J. Harris, A. R\'akos, and G. M. Sch\"utz, Europhys. Lett. \textbf{75}, 227 (2006); A. Puglisi, L. Rondoni, and A. Vulpiani, J. Stat. Mech. P08010 (2006); P. Visco, J. Stat. Mech. P06006 (2006).

\bibitem{west1982} B. J. West and V. Seshadri, Physica A \textbf{113}, 203 (1982). 

\bibitem{uchaikin1999} V. V. Uchaikin and V. M. Zolotarev, \textit{Chance and Stability: Stable Distributions and their Applications} (VSP, Utrecht, 1999).

\bibitem{caceres1997}  M. O. C\'aceres and A. A. Budini, J. Phys. A: Math. Gen. \textbf{30}, 8427 (1997); M. O. C\'aceres, J. Phys. A: Math. Gen. \textbf{32}, 6009 (1999).

\bibitem{ellis1985} R. S. Ellis, \emph{Entropy, Large Deviations, and Statistical Mechanics} (Springer, New York, 1985). 

\bibitem{note3} H. Touchette and E. G. D. Cohen, in preparation.

\bibitem{buguin2005}  A. Buguin, F. Brochard, and P.-G. de Gennes, Eur. Phys. J. E \textbf{19}, 31 (2006).

\bibitem{degennes2005} P.-G. de Gennes, J. Stat. Phys. \textbf{119}, 953 (2005).

\bibitem{note4} The Gaussian and L\'evy white noises needed for the experiment of \cite{buguin2005} can be generated by a computer and imposed externally using transducers.

\bibitem{aumaitre2001} S. Auma\^itre, S. Fauve, S. McNamara, and P. Poggi, Eur. Phys. J. B \textbf{19}, 449 (2001); P. Visco, A. Puglisi, A. Barrat, E. Trizac, and F. van Wijland, Europhys. Lett. \textbf{72}, 55 (2005); A. Puglisi, P. Visco, A. Barrat, E. Trizac, and F. van Wijland, Phys. Rev. Lett. \textbf{95}, 110202 (2005); P. Visco, A. Puglisi, A. Barrat, E. Trizac, and F. van Wijland, J. Stat. Phys. \textbf{125}, 533 (2006).

\bibitem{bonetto2006b} F. Bonetto, G. Gallavotti, A. Giuliani, and F. Zamponi, J. Stat. Mech. P05009 (2006).

\end{thebibliography}
\end{document}